\documentclass{cmspaper}
\begin{document}


\begin{titlepage}

  \conferencereport{2003/052}
  \date{1 November 2003}

  \title{The CMS Silicon Strip Tracker: \\
System Tests and Test Beam Results}

  \begin{Authlist}
    K. KLEIN
       \Instfoot{cern}{I.\,Physikalisches Institut B, RWTH Aachen \\
52074 Aachen, Germany\\ 
E-mail: Katja.Klein@physik.rwth-aachen.de}
  \end{Authlist}

\collaboration{On behalf of the CMS Tracker Collaboration}

  \begin{abstract}
With a total area of 210 squaremeters and about 15000 single 
silicon modules the silicon strip tracker of the CMS experiment at the LHC 
will be the largest silicon strip detector ever built. 
While the performance of the 
individual mechanical and electronic components has already 
been tested extensively, their 
interplay in larger integrated substructures also has to be studied before 
mass production can be launched, in order 
to ensure the envisaged performance of the overall system. This is the main 
purpose of the system tests, during which hardware components as final as 
possible are being integrated into substructures of the tracker subsystems. 
System tests are currently 
being carried out for all subsystems of the tracker. 
In addition, silicon modules and electronic components have been 
operated and studied in a particle beam environment. In this report 
results from the CMS silicon tracker system tests and a test beam experiment 
at CERN are presented.
  \end{abstract} 

\conference{Presented at the {\it 8th ICATPP Conference}, Como, Italy, 
October 6-10, 2003}
  
\end{titlepage}

\setcounter{page}{2}


\section{The CMS Silicon Strip Tracker}
\begin{figure}[ht]
  \begin{center}
    \resizebox{10cm}{!}{\includegraphics{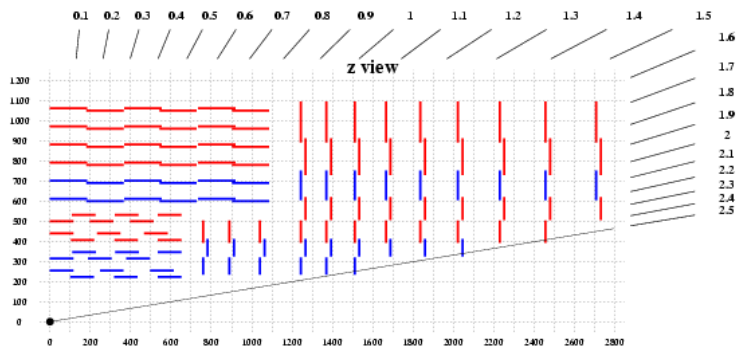}}
\caption{The CMS silicon strip tracker. One quarter of the detector is shown 
in the longitudinal view. Each line represents a silicon module.}
    \label{fig:cmstracker}
  \end{center}
\end{figure}

The CMS silicon strip tracker is divided into four subsystems: the Tracker 
Inner Barrel and Inner Disks (TIB and TID), the Tracker Outer 
Barrel (TOB) and the Tracker End Caps (TEC). The modularity of the system can 
be seen in Fig.~\ref{fig:cmstracker}, where one quarter of the detector is shown 
in the longitudinal view. The total tracker will be cooled 
to an operating temperature of --$10^{\circ}$\,C. A detailed description of the 
layout of the silicon strip tracker is available in Ref.~\cite{cms} and 
references therein.\\
Silicon modules mounted within a radial distance of 60\,cm from the beam 
line have 320\,$\mu$m thick sensors, while the sensors of all outer modules 
have a thickness of 500\,$\mu$m. Single- and 
double-sided modules are used, the latter being made of two single-sided 
modules mounted back-to-back with a stereo angle of 100\,mrad. The sensor 
design is described in Ref.~\cite{sensors}. \\
The TIB consists of four cylindrical layers. Each layer is constructed out 
of two carbon fiber (CF) half-shells per beam (z) direction. Strings carrying 
three thin modules are mounted inside and outside of the layer surfaces. \\
The TOB is composed of six cylindrical layers. The basic structure of the 
TOB is a rod: a CF support frame, which carries either three 
double-sided (layers 1-2) or three single-sided (layers 3-6) thick 
modules on each side. \\
Finally, each of the two endcaps of the TEC consists of nine CF 
disks. On each disk 16 petals, wedge shaped CF support plates which carry up 
to 28 modules arranged in seven radial rings, are mounted. \\
The readout is based on the APV25 chip\cite{apv25} built in radiation hard 
0.25\,$\mu$m CMOS technology. This 128 channels chip implements a 
charge-sensitive amplifier, a shaper and a 192 cells pipeline (3.2\,$\mu$s 
long). Two 
operation modes can be chosen: in peak mode only one data sample is 
processed, while in deconvolution mode three consecutive samples 
are summed with weights. This leads to a much shorter pulse and thus to 
correct bunch crossing identification in the high luminosity phase of the LHC.
The signals of two chips are multiplexed onto one data line and converted 
to optical ones in Analog Opto-Hybrids (AOHs)\cite{opto}. The data are then 
transmitted to the control room, where VMEbus readout boards called Front End 
Drivers\cite{fed} (FEDs) provide opto-electrical conversion, 
digitization and zero-suppression. \\
The monitoring and control is handled by Front End Controller (FEC) 
VMEbus cards, which communicate via a digital optical 
link\cite{opto} in a token ring protocol 
with dedicated Communication and Control Units (CCU25)\cite{ccu25} mounted 
on the string/rod/petal motherboards. These chips distribute the control 
signals 
to the addressed modules, while trigger and clock signals are propagated to 
Phase Locked Loop (PLL) chips on the front-end hybrids.

\section{Results from System Tests and the May 2003 Beam Test}
System tests are currently 
being carried out for all subsystems of the tracker: for the TOB at 
CERN, for the TIB/TID in Florence and Pisa and for the TEC in Aachen and Lyon. 
Both electrical behaviour, with emphasis on the noise and signal-to-noise 
performance, and the cooling performance are being 
studied and the design is qualified or optimized, if necessary. \\
For the TIB and TEC the most complete system tests up to now 
have been realized in a test beam environment at CERN during May 2003. 
The X5 beam 
in the CERN West Area provided muons and/or pions ($p\,$=\,120\,GeV/$c$ for 
pions). The beam had a LHC-like time structure, with about 3\,nsec long 
particle bunches, spaced by 25\,nsec time periods. \\
The main difference between test beam and system test setups 
and the final CMS readout and control chains is that currently PCI mezzanine 
cards (PMC) are used for readout and control (PMCFED\cite{pmcfed} and PMCFEC) 
instead of the final VMEbus cards. These PMC have no implementation of 
optical conversion, thus additional opto-electrical converters are 
necessary. \\
In the following the setups of the TEC and TIB beam tests and the TOB 
laboratory system test, along with first preliminary results, are described.

\subsection{Test Beam Data Acquisition}
In the test beam the most recent DAQ software, based on the 
XDAQ\cite{xdaq} framework, was used. For the first time, a prototype of 
the final run control\cite{runcontrol} was available, and an online
 monitoring programme provided immediate feedback on the performance. For 
each subsystem (TIB, TOB, TEC) optical transmission 
of data as well as timing and control signals between the control room and 
the beam area was realized. Each subsystem used one PC with a PMCFEC and a 
second PC housing two or three PMCFEDs and a Trigger Sequencer Card, 
which distributed the particle trigger and the clock from the 
TTC system\cite{ttc} to the FEC and the FEDs. \\
The commissioning of the individual subsystems was finished within about two 
hours. This included the tuning of 
the optimal FED sampling point, the adjustment of the timing difference 
between individual channels due to their different positions in the trigger 
distribution path, the optimization of AOH parameters (adjustment of the 
laser diode gain and bias current, to be repeated for each temperature 
change) and finally trigger latency and PLL delay scans to find the physics 
signal (the sampling point which gives the highest signal to noise). 
Automatized procedures are implemented in the software for all these tasks. 
Mostly the TIB, TEC and TOB subsystems were read out 
independently of each other, but finally the TIB and 
TOB DAQ systems were merged and the two subdetectors read out coherently 
like a single detector after only a few hours of commissioning. This shows the 
scalability and commissioning capability of the DAQ software.

\subsection{The Tracker End Cap Beam Test}
\begin{figure}[Ht]
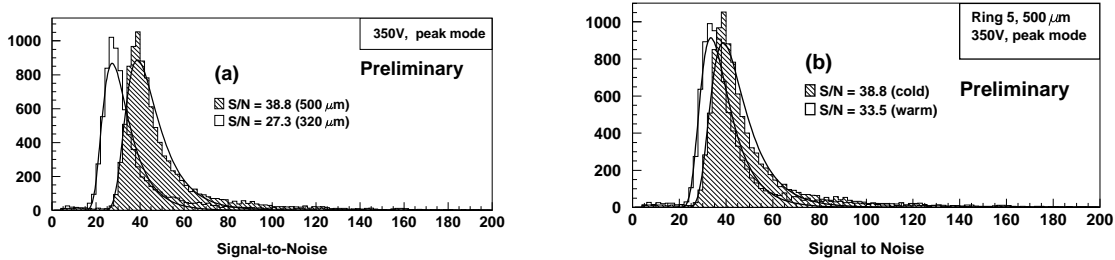

  \begin{center}
  \resizebox{7.cm}{!}{\includegraphics{KKlein_fig2.epsi}}\hspace{1cm}
  \resizebox{7.cm}{!}{\includegraphics{KKlein_fig3.epsi}}
  \caption{Signal-to-noise distributions as measured in the TEC beam test, 
(a) for a ring four module (thin sensor, blank histogram) and a ring five 
module (thick sensors, filled histogram) at $0^{\circ}$C; (b) for the 
ring five module at $0^{\circ}$C (filled histogram) and 
room temperature (blank histogram). All runs were taken in peak mode.}
  \label{fig:TEC} 
  \end{center}
\end{figure}

\begin{figure}[ht]
  \begin{center}
    \resizebox{8cm}{!}{\includegraphics{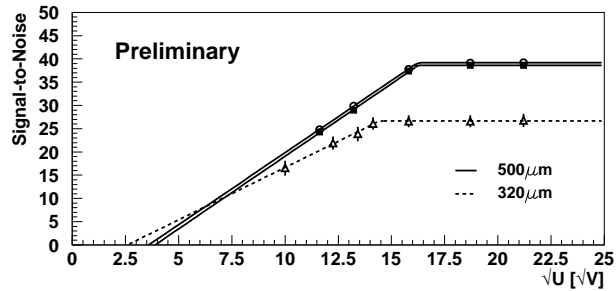}}
\caption{Signal-to-noise ratio versus the 
square root of the bias voltage for a TEC ring four 
module (dashed line) and a ring five module (solid lines, both sensors are 
shown).}
    \label{fig:biasscan}
  \end{center}
\end{figure}

For the first time a prototype of a 
TEC petal, equipped with nine modules (four thin single-sided modules on 
ring four, four thick single-sided modules on ring six, plus one thick 
double-sided module on ring five), was studied in a test beam. Twelve 
front-end hybrids plus AOHs were 
distributed on the remaining positions. The petal was cooled via its own 
cooling pipe system and was kept inside a thermally and electrically 
isolated passive cooling box, flushed with dry 
nitrogen. Temperature and humidity inside the box were monitored and an 
interlock on the low voltage was implemented. Floating power supplies 
(not of the final design) were used for low voltage and also for high voltage 
to bias the detectors. 
Both low and high voltages were transmitted to the petal via 
45\,m long cables of the final design.  \\
The system showed excellent performance in terms of the signal-to-noise ratio 
(S/N). The S/N distributions of all modules have been studied in peak mode 
running for a bias voltage of 350\,V and an operating temperature of 
$0^{\circ}$\,C. In Fig.~\ref{fig:TEC}\,(a) two examples are shown: 
S/N Landau peaks of 27 and 39 are found for a thin and 
a thick module, respectively. 
In Fig.~\ref{fig:TEC}\,(b) the S/N of the same thick (ring five) 
module is compared for running at $0^{\circ}$\,C and at room 
temperature, under otherwise identical conditions. The 
S/N is increased significantly when the module is operated at $0^{\circ}$\,C, 
compared to operation at room temperature, where a S/N of only 33.5 is 
measured.\\
Figure~\ref{fig:biasscan} shows the results of bias voltage scans. 
For a ring four module a plateau in S/N is reached for a 
voltage of 190\,V. A double-sided ring five module consists of two 
single-sided modules, each with two daisy-chained wafers, mounted with a 
stereo angle. For the two sensors of one single-sided 
ring five module the plateau voltages are 264\,V and 265\,V. It was 
possible to distinguish between the two sensors since the stereo angle 
was exploited to calculate the radial coordinate along the 
strip direction. In this way the clusters can be assigned to the individual 
sensors. The plateau voltage is higher than the depletion voltage, since 
charge collection is incomplete without significant overdepletion.
High plateau voltages at the start ensure that after type inversion due to 
irradiation only moderate bias voltages must be applied to maintain full 
efficiency.\\

\subsection{The Tracker Inner Barrel Beam Test}
\begin{figure}[ht]
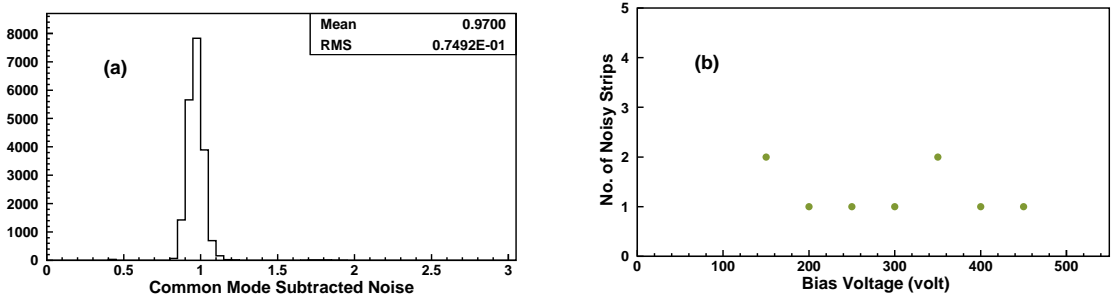

  \begin{center}
    \resizebox{6.8cm}{!}{\includegraphics{KKlein_fig5.epsi}}\hspace{1cm}
    \resizebox{6.8cm}{!}{\includegraphics{KKlein_fig6.epsi}}
\caption{Noise measurements with a TIB module in the test beam: (a) 
common mode subtracted noise; (b) the bias voltage dependence of the 
number of noisy strips.}
    \label{fig:tibnoise}
  \end{center}
\end{figure}

The TIB test beam setup consisted of a part of a half-shell of 
layer three with four strings. Two of these strings were equipped with three 
single-sided thin modules each, while the CCU25 and the mothercable were 
mounted for all four strings. The 
TIB setup was thermally stabilized at room temperature. Temperature and 
humidity were monitored and interfaced to an interlock system. Two different 
protoypes of the final control room power supplies (from 
CAEN and LABEN), supplying both low and high voltage, 
as well as a 125\,m long low inductance power cable of the final design were 
studied for their noise behaviour, and exhibited excellent performance. \\
The noise of the system was found to be very low. The common mode subtracted 
noise of a typical TIB module is shown in Fig.~\ref{fig:tibnoise}\,(a). 
The mean common mode subtracted noise is only 0.97 ADC counts. In Fig.~\ref{fig:tibnoise}\,(b) the 
dependence of the number of noisy strips of one module on the bias 
voltage is shown. The number of noisy strips is very small and stable. \\
The signal pulse shape, measured with a muon beam, has been reconstructed in 
peak and deconvolution mode. While in peak mode a (slightly adjustable) 
peaking time of about 55\,nsec is found, in deconvolution mode the pulse 
is much sharper and the peaking time is below 20\,nsec. This is known to be 
achieved at the cost of a lower S/N. However, with a typical 
Landau peak of 18 (for 300\,V bias voltage), the S/N in deconvolution mode is 
still sufficiently high. In peak mode a S/N of about 26 is measured. 

\subsection{The Tracker Outer Barrel System Test}
\begin{figure}[ht]
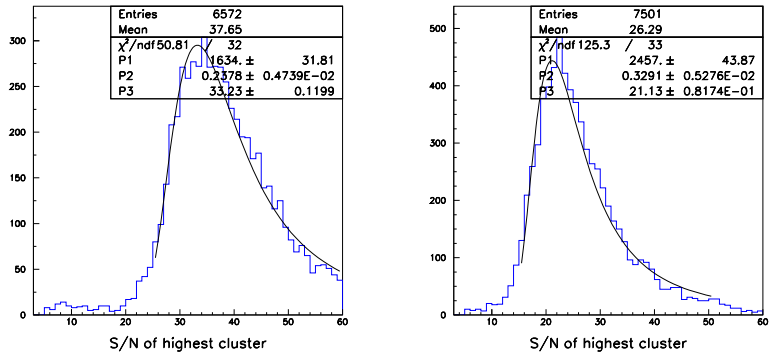

  \begin{center}
    \resizebox{4.5cm}{!}{\includegraphics{KKlein_fig7.epsi}}\hspace{1cm}
    \resizebox{4.5cm}{!}{\includegraphics{KKlein_fig8.epsi}}
\caption{Signal-to-noise ratio measured on a TOB module using a 
Ruthenium $\beta$-source, in peak mode (left) and deconvolution mode 
(right).}
    \label{fig:TOB}
  \end{center}
\end{figure}

The system test of the TOB is in a very advanced state. The system test 
of a single-sided rod is finished and the design has been validated. 
Currently a double-sided rod equipped with twelve $r$--$\phi$ modules is under 
test at CERN, and a CMS Note summarizing the results is in preparation. \\
The noise performance of the rod was tested extensively. A comparison 
between a single module setup, consisting of a bare module, and 
the rod setup equipped with the same module shows compatible noise and 
common mode both in peak and deconvolution mode. \\ 
To test the TOB modules with real particles, a cosmic test stand has been 
realized. With this setup a S/N of 26 is measured in deconvolution mode. \\
For faster measurements, important during the mass production phase, the 
modules are exposed to a $^{106}_{44}$Ru $\beta$-source, which provides 
electrons with a maximal energy of 3.5\,MeV and a trigger rate of 500\,Hz. 
For electrons, the S/N is typically 33 in peak mode 
and 21 in deconvolution mode (Fig.~\ref{fig:TOB}). The 
difference in S/N for cosmic muons and 
electrons is mainly due to the different mean path lengths in the silicon. 
The double-sided modules have been exploited 
to calculate the hit efficiency, which is found to be as high as 
99.8\,\%. First tracking and alignment studies at the overlap of two 
double-sided modules were already carried out. \\

\section{Conclusions}
The system tests of the TEC, TIB and TOB subdetectors of the CMS silicon 
strip tracker are in an advanced state. Increasingly more complex 
substructures are being integrated and studied in laboratory system 
tests as well as in test beam experiments. Up to now the design has been 
proven to work very well, exhibiting low noise and excellent signal-to-noise 
ratio performance. In the TEC system test a full petal will be 
integrated until the end of the 
year, while for the TIB the next step is the integration of four full 
single-sided 
strings. Mass production of silicon strip modules has started, and the first 
fully equipped substructures will be installed on the disk and barrel 
structures in the first half of 2004.


\end{document}